\documentstyle[12pt,epsfig]{article}

\newlength{\dinwidth}
\newlength{\dinmargin}
\newcommand\bea{\begin{eqnarray}}
\newcommand\eea{\end{eqnarray}}

\newcommand{\be}{\begin{equation}}
\newcommand{\ee}{\end{equation}}
\newcommand{\bc}{\begin{center}}
\newcommand{\ec}{\end{center}}

\begin{document}
\thispagestyle{empty}
\addtocounter{page}{-1}
\vskip-0.35cm
\begin{flushright}
\end{flushright}
\vspace*{0.2cm}
\centerline{\Large \bf Fixed points and FLRW cosmologies:}
\centerline{\Large \bf Flat case}
\vspace*{0.5cm}
\centerline{\bf Adel Awad \footnote{\tt aawad@zewailcity.edu.eg}}

\vspace*{0.4cm}
\centerline{\it Centre for Theoretical Physics, Zewail City of Science and Technology, }
\vspace*{0.2cm}
\centerline{\it Sheikh Zayed, 12588, Giza, Egypt.}
\vspace*{0.2cm}
\centerline{\it Department of Physics, Faculty of Science, Ain Shams University, Cairo, 11566, EGYPT}
\vspace*{0.2cm}

\vspace*{0.5cm} \centerline{\bf Abstract} \vspace*{0.3cm}
 We use phase space method to study possible consequences of fixed points in flat FLRW models. One of these consequences is that a fluid with a finite sound speed, or a differentiable pressure, reaches a fixed point in an infinite time and has no finite-time singularities of types I, II and III described in {\tt hep-th/0501025}. It is impossible for such a fluid to cross the phantom divide in a finite time. We show that a divergent $dp/dH$, or a speed of sound is necessary but not sufficient condition for phantom crossing. We use pressure properties, such as asymptotic behavior and fixed points, to qualitatively describe the entire behavior of a solution in flat FLRW models. We discuss FLRW models with bulk viscosity $\eta \sim \rho^r$, in particular, solutions for $r=1$ and $r=1/4$ cases, which can be expressed in terms of Lambert-W function. The last solution behaves either as a nonsingular phantom fluid or a unified dark fluid. Using causality and stability constraints, we show that the universe must end as a de Sitter space. Relaxing the stability constraint leads to a de Sitter universe, an empty universe, or a turnaround solution that reaches a maximum size, then recollapses.

\vspace*{0.5cm}
\baselineskip=18pt


\newpage

\section{Introduction}

Various cosmological observations \cite{Riess,wmap,SDSS,Cha-Xray} have provided us with a strong evidence for accelerating expansion of the universe. Although, the component that causes this acceleration is not known yet, the best model that fits dark energy and other cosmological data is the $\Lambda$CDM model, which is a Friedmann-Lemaître-Robertson-Walker (FLRW) universe with a cosmological constant. Unfortunately, this model does not provide a physical picture of dark energy. In order to describe dark energy in the realm of general relativity, we need to consider some exotic fluid with an unusual equation of state that has a negative pressure and violates the strong-energy condition (see for example \cite{caldwell_review,tuener_review06}). The existence of this exotic fluid not only opens the door for reexamining the constituents of our universe but also evades some nonsingularity theorems through relaxing the strong-energy condition. This revives interest in nonsingular cosmologies, especially those describing the universe in early and late times. Having a more general class of equations of state, that violates the strong-energy condition, creates a wider class of singularities beyond that of Big Bang and Big Crunch \cite{br,barrow1,Bouhmadi-Lopez}. In FLRW models these singularities are classified as follows \cite{no-od-class}:\\
$\bullet$ Type I ("Big Rip"): $t\rightarrow t_s$, $ a \rightarrow\infty$, $\rho \rightarrow \infty$, and $|P| \rightarrow \infty$\\
$\bullet$ Type II ("Sudden"): $t\rightarrow t_s$, $ a \rightarrow a_s$, $\rho \rightarrow \rho_s$, and $|P| \rightarrow \infty$\\
$\bullet$ Type III : $t\rightarrow t_s$, $ a \rightarrow a_s$, $\rho \rightarrow \infty$, and $|P| \rightarrow \infty$\\
$\bullet$ Type IV : $t\rightarrow t_s$, $ a \rightarrow a_s$, $\rho \rightarrow 0$, and $|P| \rightarrow 0$, but higher derivative of $H$ diverges.\\
It is of interest to build cosmological models, which are free from the above singularities.

In the last decade, there have been several proposals to describe dark energy/matter or dark energy alone as a single barotropic fluid in FLRW models with various equations of state (EoS) such as; Chaplygin gas, Van der Waal, linear, and quadratic EoS\cite{KMP-2001,BTV-2002,BBS-2002,CSN-1997,CSN-1998,BDE-2005,HN-2004,AB-2006,QBB-2007,NO-2004,Kremer-2003,Capoz-2005}. Notice that, most of these models have fixed points, or de Sitter solutions.

In this article we use a phase space method to study general solutions of single fluid FLRW models with fixed points and a pressure $p(H)$, where $H$ is the hubble parameter. We discuss possible consequences of having these fixed points. Some of these consequences are; {\it (i)} if we describe our universe as a single component fluid and model the late time acceleration by a future fixed point, then the resulting cosmology does not have future-time singularities of types I, II and III described in \cite{no-od-class},  {\it (ii)} cosmologies with a future and a past fixed points are free of types I, II and III singularities, {\it (iii)} one can use a simple argument to show the phantom divide \cite{Hu-2005,Knuz-2006}, or in a single fluid FLRW models it is impossible for a physical solution to cross the phantom divide line in a finite time, and {\it (iv)} in these models, the only way to get bounce solutions is to have a nonvanishing pressure as $\rho \rightarrow 0$. The phase space method can be used to construct nonsingular late time model, in particular, unified dark fluid (UDF) and dark energy models. We use this qualitative analysis to describe the entire behavior of a FLRW cosmology with bulk viscousity $\eta \sim \rho^r$, presenting two exact solutions, with $r=1$ and $r=1/4$ expressed in terms of Lambert-W function. The last solution describes either a nonsingular phantom dark energy or a unified dark fluid model. At the end of section 4 we list possible scenarios for the future of our universe.

\section{Flowing to a Fixed Point}

In this section, we use a phase space method in flat FLRW models with fixed points to argue that a fluid with a continuous and differentiable pressure always reaches a fixed point in an infinite time and has no finite-time singularities of types I, II and III described in \cite{no-od-class}.

Let us start with a FLRW universe, with an equation of state
                    \be p=p(H),\ee
where the pressure $p(H)$ is a continuous function of the hubble parameter $H$. Using the unit convention, $8\pi G=c=1$, Einstein field equations lead to Friedmann equation and Raychaudhri equation
\be H^2={\rho \over 3}-{k \over a^2}, \label{friedmann}\ee \be\dot{H}=-H^2-{1\over 6}\,(\rho+3\,p)) \hspace{0.3in} H={\dot{a}/ a} \label{H}.\ee
For energy-momentum conservation, we obtain
\be \dot{\rho}=-3\,H\,(p+\rho) \label{one}\ee
which is not independent of Eqn. (\ref{H}).
In this work we are interested in the flat FLRW case, the spatially curved case will be discussed elsewhere \cite{curved_1}. For a flat FLRW universe, Eqn.(\ref{H}) becomes
\be  \dot{H}=-{1\over 2}(p(H)+\rho)=f(H),\ee
which can be expressed in terms of a dimensionless hubble parameter and time $h=H/H^*$ and $\tau=c\,H^*\,t$, as follows
\be \label{1dphase} { dh \over d\tau}= F(h).\ee
Where $H^*$ is a parameter that depends on the equation of state parameters and c is some number.
A solution, $h(\tau)$ of Eqn. (\ref{1dphase}) is subject to an initial condition $h(0)=h_0$.

Motivated by the fact that many of the proposed models for dark energy and unified dark matter/energy models do have fixed points, we assume the existence of fixed points for Eqn.(\ref{1dphase}), which are the zeros of the function $F(h)=-1/2\,(\rho+p)$. Let us call the zeros of $F(h)$, $h_1, h_2,...$, where $h_1<h_2<h_3..$.

\begin{figure}[htp]
 \centerline
  {\includegraphics[angle=-90,width=100mm]{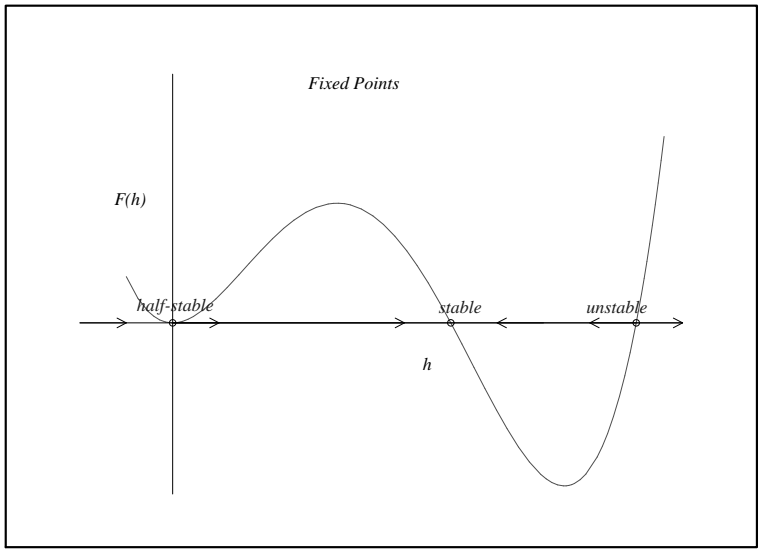}}
 \caption{\footnotesize Types of fixed points}
 \label{fig-1}
 \end{figure}

Fixed points are classified according to their stability as follows; stable, unstable, or half-stable, depending on the sign of their tangents as shown in Figure (\ref{fig-1})\footnote{Here I am following the terminology of Strogatz \cite{strogatz}.}. Unstable fixed points are represented by arrows emanating out of them and stable points are represented by arrows pointing toward them as shown in Figure (\ref{fig-1}). Half-stable points are stable from one side and unstable from the other side or vice versa.

 The arrows determine how the solution develops with time. Notice that, a fixed point satisfies equation (\ref{1dphase}), i.e., $h(\tau)=h_1$, therefore, it is a solution. This constant solution is nothing but a de Sitter space. If the system started from an initial value, $h_0=h_1$, i.e., at a fixed point, then it will remain at this point forever. But when the value of $h_0$ is close to that of $h_1$, the solution $h(\tau)$ develops towards $h_1$, if $h_1$ is a stable fixed point, or away from it, if the point is unstable. This technique has been used in literature to study particular equations of state and their solutions (e.g., \cite{BDE-2005}), but here we try to keep our discussion general assuming a pressure $p(H)$.

Using this tool, one can determine how a solution behaves upon knowing the nature of these fixed points. For example, if we start with a value for $h_0$ between the stable and the unstable fixed points in Figure (\ref{fig-1}), then the solution will develop to the left, i.e., will take values with smaller $h$ till it reaches the stable point. Therefore, the solution in this case interpolates between two de Sitter spaces. If $h_0$ takes values larger than the unstable fixed point in Figure(\ref{fig-1}), then the solution develops to the right and might reach a singularity if the time to reach $h\rightarrow \infty$ is finite.

Fixed points and the asymptotic behavior of $F(h)$ enable us to predict the behavior of the system without knowing the form of the solution. But, in order to have a reliable qualitative description of a solution we have to know how long it takes to reach a fixed point or a singularity, i.e., a point where $h\rightarrow \infty$. Using the argument below we will see that the continuity and differentiability of the pressure, $p(h)$ determine if the time to reach a fixed point is finite or infinite. Also, in the coming section we use the asymptotic behavior of $F(h)$ to determine the time taken by a solution to reach a point where $F(h)$ is diverging. We are going to use pressure properties to describe the behavior of a solution qualitatively without the need for exact or approximate solutions.\\
Here we argue that for a flat FLRW fluid with a pressure $p(h)$, which satisfies;\\
i) $F(h)$ is continoues and differentiable, and\\
ii) $F(h)$ has a future fixed point $h_1$, i.e., $h_0 < h_1$ if $F(h_0)>0$, (or $h_0 > h_1$ if $F(h_0)<0$), \\
there exists a unique solution $h(\tau)$ which is defined for times $\tau >0$, and has no future finite-time singularities of types I, II and III. \\
Notice that, the first assumption is needed to ensure the existence of a unique local solution around some initial value $h_0$ by the existence-uniqueness theorem. In addition, it leaves the solution free from type-II singularities, which might contradict casuality since the sound speed $d p/d\rho=c_s^2$ will diverge as well. The second assumption ensures the extendibility of this local solution to all values $\tau >0$ while keeping the hubble parameter bounded.\\

Let us start our argument by integrating equation (\ref{1dphase}), assuming an initial value $h(0)=h_0$, then one obtains
\be G(h)=\int_{h_0}^{h} {dy \over F(y)}=\tau\ee

Since $F(h_0) \neq 0$, its either $F(h_0) >0$ or $F(h_0) <0$, let us choose $F(h)>0$, and $h_0<h_1$ from ii). As a result $G(h)$ is a monotone near $h_0$. For any solution $\phi(\tau)$, we have
\be G(\phi(\tau))=\tau.\ee
Since $G(h)$ is a monotone near $h_0$, the above relation can be inverted
\be G(\tau)^{-1}=\phi(\tau),\ee
where $G^{-1}$ is the inverse map of $G$. This is a local solution by construction (unique since $F(h)$ is differentiable by the existence-uniqueness theorem) that can be extended by looking for a maximal interval in which $G(h)$ is a monotonic function. Since $F(h_0)>0$ and remains positive for the values $h_0<h<h_1$, then $G(h)$ is a monotone in this interval. This imply that the maximum interval in which we can extend the solution is [$h_0,h_1$].
The solution can be defined for all values $\tau>0$ if \be \tau_+ =\int_{h_0}^{h_1} {dh \over F(h)} =\infty. \ee
Now, let us show that $\tau_{+}=\infty$, when i) and ii) are given.
Since $F(h)$ is differentiable, the slope of the tangent at any point is finite. Let us choose a number $M<F'(h),\forall h\in [h_0,h_1]$. The existence of $M$ enables us to define a linear function Y(h) such that \be Y(h)\geq F(h),\label{lin-fn}\ee where $Y(h)=M\, (h-h_1)$.
This leads to
\be \int_{h_0}^{h_1} {dh \over F(h)} \geq \int_{h_0}^{h_1} {dh \over F'(h_{min})\,(h-h_1)} =\infty , \ee
therefore,
\be \tau=\int_{h_0}^{h_1} {dh \over F(h)} =\infty. \label{inf-time}\ee

We have shown that for the first-order system of Eqn.(\ref{1dphase}) there exists a unique solution, $h(\tau)$ defined for times $\tau>0$ if the above two assumptions are satisfied. Also, it takes the solution an infinite time to reach the future fixed point $h_1$.
It is clear that the solution is bounded, i.e., $h(\tau) \in [h_0,h_1]$ for times $\tau >0$. Therefore, the density, $\rho(\tau)$ is bounded for times $\tau >0$, as a result, there is no future singularities of type I and III. The pressure $p=-2\,F(h)-3\,h^2\,{H^*}^2$ is bounded too in this interval, since $F(h)$ and $h$ are bounded\footnote{Since any continuous function on a closed interval is bounded}, which means, no future singularities of type II in these spacetimes. As a result this class of FLRW solutions are free from future finite-time singularities of type I, II and III .\\

The above argument can be generalized by relaxing the differentiability condition, in this case we have one of the following;\\
a) $F'(h)$ has a jump discontinuity: $F(h)$ is not continuous, but its left and right derivatives\footnote{ Left derivative is defined as ${d_{-}F (a)\over dh}=\lim_{h\rightarrow a^-} {F(h)-F(a) \over h-a}$ and the right derivative is defined in a similar way.} are finite for $h\in[h_0,h_1]$. In this case, one can still construct the linear function $Y(h)$, which can be used to show that the time to reach $h_1$ is always infinite and the solution has no future finite-time singularities of types, I, II, and III.\\
b) $F(h)$ has infinite discontinuity: $F(h)$ is not continuous and at least the left or right derivative of $F(h)$ is infinite for some $h\in[h_0,h_1]$. This case is not physical, since the divergence of $F'(h)$ leads to a divergent speed of sound, $dp/d\rho=c_s^2$, which has to be less than unity for the model to be causal. We are going to discuss this issue in more details in the coming section when we discuss phantom crossing, but through out this paper we are going to assume no infinite discontinuities in $F'(h)$.

\section{Consequences of Fixed Points}

Here we discuss consequences of fixed points and how can we use this dynamical method to describe the entire behavior of a single fluid in flat FLRW without knowing the form of the solution.

\subsection{Direct Consequences}

Consequences of using phase space method to study fixed points can be listed as follows;\\

\begin{figure}[htp]
 \centering
  {\includegraphics[angle=-90,width=82mm]{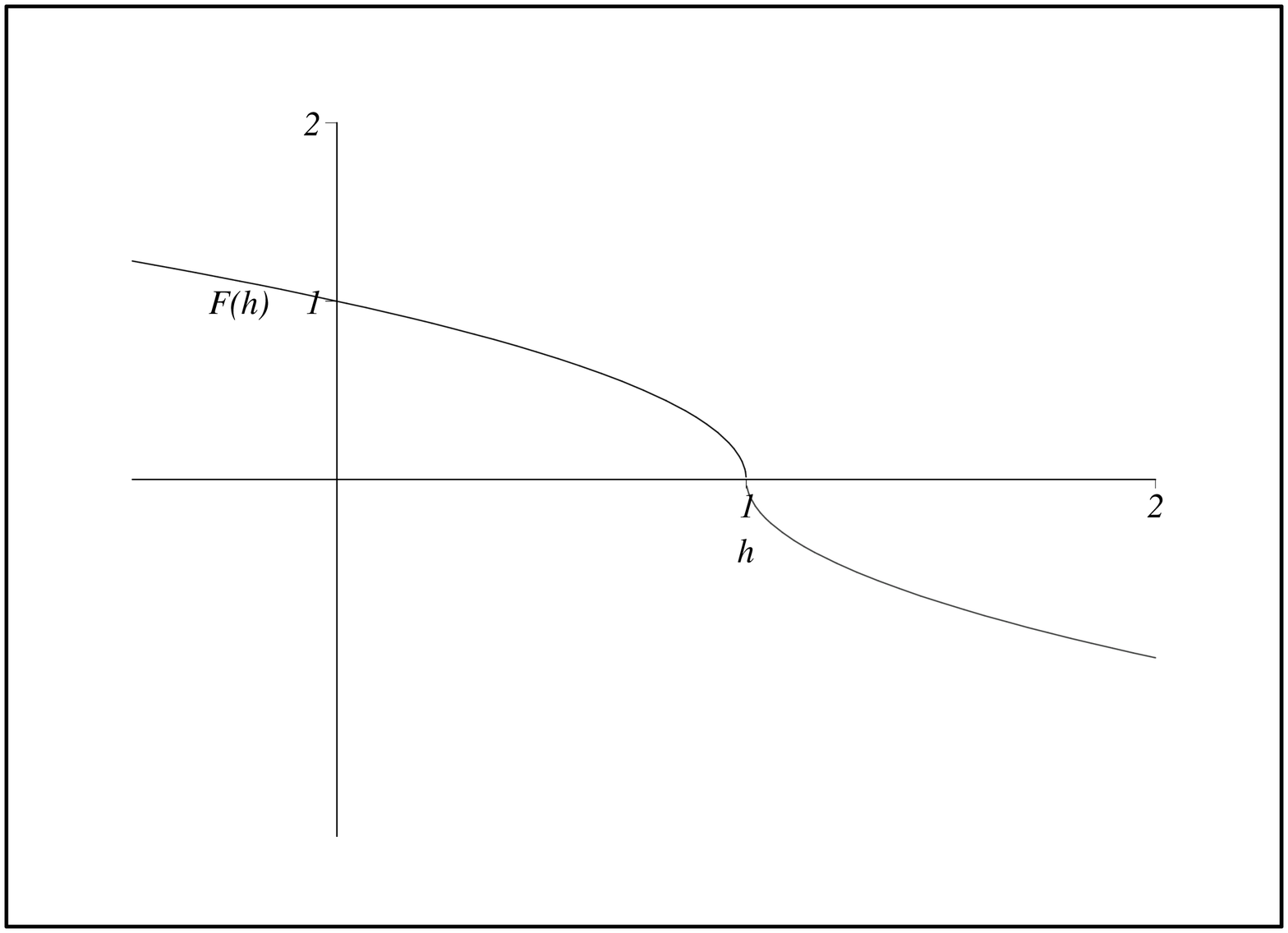}}{ \includegraphics[angle=-90,width=82mm]{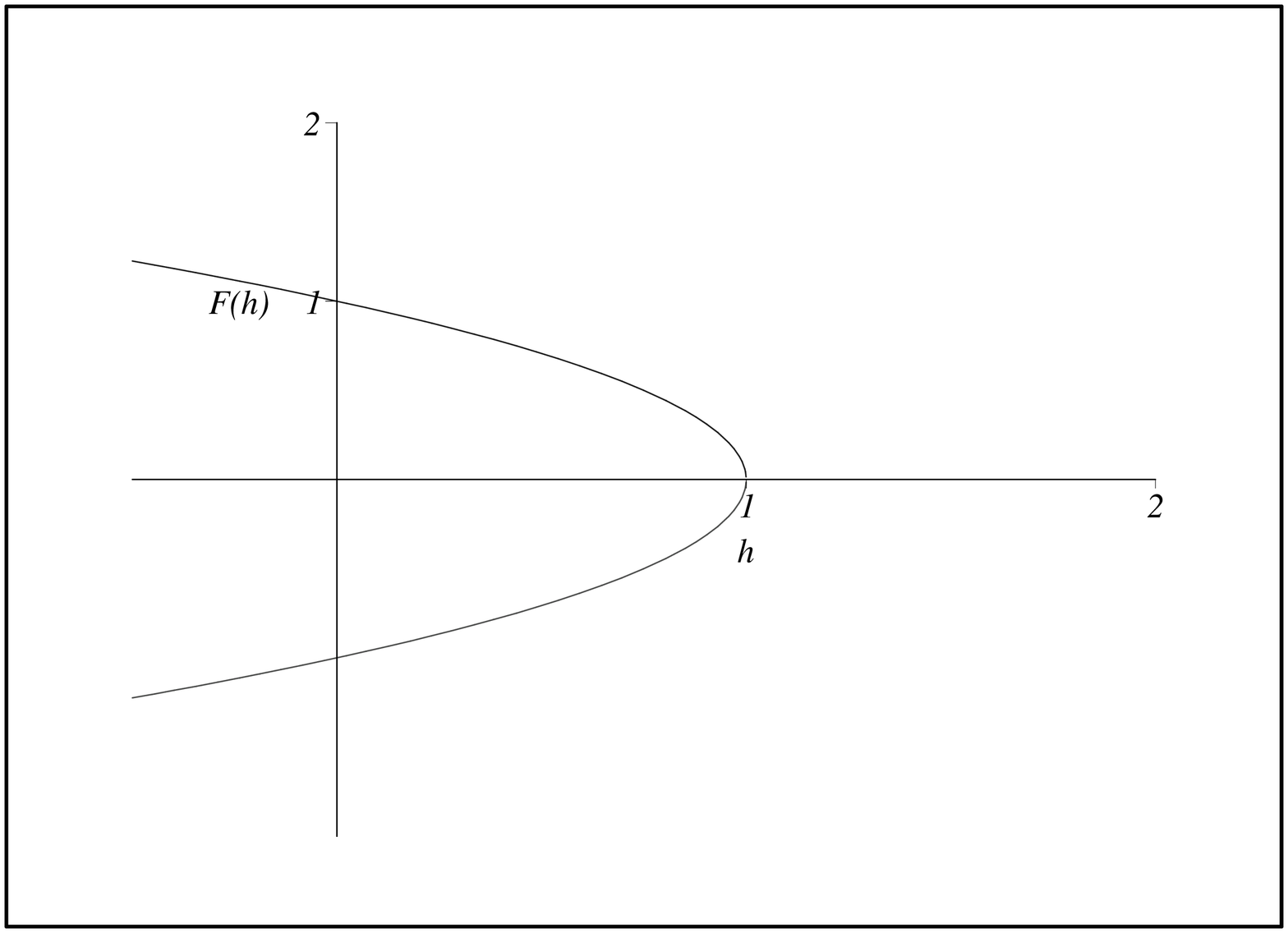}}
 \caption{\footnotesize {\bf a)} $F(h)=(1-h)^{1/2}$, for $h\leq 1$, and $F(h)=-(h-1)^{1/2}$, for $h> 1$. {\bf b)} $F(h)=\pm \,(1-h)^{1/2}$, for  $h\leq 1$.}
 \label{fig-3}
 \end{figure}

{\it i)} If the late-time behavior of our universe is described by a single fluid, as in unified dark fluid models, and the late-time acceleration is developing towards a de Sitter universe, then there is no future-time singularity of types I, II and III in this solution. Since the pressure $p(h)$, is a differentiable function of $h$, or at most has finite discontinuities, it takes the universe infinite time to reach the de Sitter space.\\
{\it ii)} If we combine a future fixed point with a past fixed point, by applying the argument in section 2 we get a nonsingular solution, which is free from types I, II and III singularities. The time taken by the solution to reach $h_1$ or to come from $h_2$ starting from an initial value $h_0$ is infinite. In addition, the hubble parameter $h(\tau)$ interpolates between $h_2$ and $h_1$ in a monotonic manner.\\
{\it iii)} As one might notice, the solutions we have so far, still, admit a weaker type of singularity, namely, type-IV in the classification given in \cite{no-od-class}. But even these weaker singularities can be avoided by requiring $p(h)$ to be a smooth function, i.e., a $C^{\infty}$ function. One can show that as follows; The n-time derivative of $h(\tau)$ can be written as;
\be h^{(n)}= \left(F(h)\,{d \over dh}\right)^{n-1}\, F(h).\ee
which means, unless $F(h)$ or one of its derivatives (up to the $(n-1)$th derivative) is divergent, $h^{(n)}$ is always bounded in $[h_2,h_1]$. This leads to the conclusion: In a flat FLRW universe, if a) $p(h)$ is a smooth function, i.e., arbitrarily differentiable, and b) $F(h)$ has a future and a past fixed point, then the spacetime is free from singularities of types, I, II, III and IV \cite{no-od-class} when $h_0\in[h_2,h_1]$.\\
{\it iv)} A single fluid in flat FLRW with a pressure $p(h)$ admits bouncing solutions only if there are no fixed points between $h=0$ and $h=h_1$, i.e., $F(0)\neq 0$. These solutions either have a bounce or turnaround at $h=0$ depending on sign of $F(0)$. At $h=0$, $\ddot{a}/a=F(0)$, therefore, if $F(0)>0$ it will be a bounce and for $F(0)<0$ it will be a turnaround.\\
{\it v)} Another consequence of this phase space method is a simple and transparent way to show the no-go theorem of phantom divide \cite{Hu-2005,Knuz-2006} (see also, \cite{No-Od-phantom,quintom}), which can be stated as follows; In a single fluid FLRW cosmology it is impossible for a causal solution to go from a region where $\omega(h)<-1$ to another with $\omega(h)>-1$. In other words, a solution in one of the mentioned regions has no access to the other region. Let us explain this in more details using the analysis we have in section 2.
First, let us assume that the pressure $p(h)$ is differentiable. In this case, if a solution approaches a fixed point, where $w(h)=-1$, starting from a region where $w(h)<-1$, then, it will spend an infinite time to reach it, as a result it will never cross it. If $p(h)$ is not differentiable, then $p'(h)$ has either a finite or an infinite discontinuity. If the discontinuity is finite, the time to reach the crossing point or the fixed point is infinite, as we showed in the previous section, therefore, the crossing will not occur. If $p'(h)$ has an infinite discontinuity, a solution will reach $w(h)=-1$ in a finite time, but in this case the solution is not causal. Although, the solution is not causal, it is not clear if it is going to cross the phantom divide or not. To further investigate this case, let us list the conditions on $F(h)$: \\i) $\lim_{h\rightarrow h_1} F(h) = 0$, \\ii)  $\lim_{h\rightarrow h_1} dF(h)/dh = \pm \infty$ and \\iii) $\tau = \int_{h_0}^{h_1} 1/F(h) < \infty$.\\
A class of functions which satisfy the above conditions is $ F(h) =F_0\, (h_1-h)^{s}$, where $0<s<1$ and $F_0>0$. A solution is given by the following expression;
\bea h(\tau)&&=h_1-\left[(h_1-h_0)^{1-s}-F_0\,(1-s)\,\tau\right]^{1 \over 1-s}, \hspace{.5in}  \tau\leq \tau^*\nonumber\\
&&=h_1 \hspace{3.04in}  \tau > \tau^* \label{dis_sol}\eea
where, $h(0)=h_0$ and $\tau^*={(h_1-h_0)^{1-s} /F_0\,(1-s)}$. It is clear from Eqn. (\ref{dis_sol}) that the solution stays at the fixed point for $\tau \geq \tau^*$. Also, notice that $h_1$ is a stable fixed point. One can see that by assuming a small perturbation away from $h_1$, i.e., $h(\tau)=h_1+\delta(\tau)$, it leads to \be \delta(\tau)= -\left[(h_1-h_0)^{1-s}-F_0\,(1-s)\,\tau\right]^{1 \over 1-s}.\ee
As a result, if we extend the definition of $F(h)$ for $h>h_1$, e.g., $F(h)=-F_0\,(h-h_1)^s$ as shown in Figure (\ref{fig-3})-a), we will not have a phantom crossing. But if we define $F(h) =\pm F_0\,(h_1-h)^s$, i.e., a double valued function as in Figure (\ref{fig-3})-b), the solution can cross the phantom divide in a finite time. Therefore, in addition to its infinite discontinuity $F'(h)$, need to be a double valued function to have a phantom crossing solution. Our conclusion is that a general causal solution of any flat FLRW model, with a continuous pressure $p(H)$, can not cross the phantom divide line in a finite time. It is clear that infinite discontinuity of $F(h)$ is necessary but not sufficient for a phantom crossing.
\subsection{Describing a Solution Qualitatively}

As we mentioned earlier, to have a complete qualitative description of a general solution in flat FLRW cosmology we need to know the fixed points as well as the asymptotic behavior of $F(h)$. The later property enables us to determine the time to reach a point where $F(h)\rightarrow \pm\infty$, starting from some initial value $h(0)=h_0$. Let us first show the relation between the asymptotic behaviors of $F(h)$ and finite-time singularities.\\
{\bf Asymptotic Behavior of $F(h)$ and Singularities:}
The dimensionless Hubble parameter $h$, in a flat FLRW cosmology is controlled by a one-dimensional phase space evolution function $F(h)$. In this dynamical system a solution develops towards either a fixed point, where $F(h) \rightarrow 0$, or a point where $F(h) \rightarrow \pm\infty$. A solution approaching a point where $F(h) \rightarrow \pm\infty$ does not necessarily mean that the it has a finite-time singularity. It is crucial to know how fast $F(h)$ reaches infinity. This enables us to determine if the singularity is reached in a finite time or not \cite{vis_littlerip}. Considering the integral \be \tau=\int_{h_0}^{h} {dh' \over F(h')},\ee it is easy to see that if $\lim_{h\rightarrow \pm\infty} F(h) \sim h^{s}$, where $0\leq s\leq1$, the integral diverges. As a result, the solution takes an infinite time to reach the singular point, therefore, it has no finite-time singularities. One can observe that if $F(h)$ grows as a linear function or slower \footnote{This is a known fact in the literature see for example \cite{vis_littlerip}}, as $h\rightarrow \pm\infty$, the solution will have no finite-time singularities. One can show this rigourously following the same argument in section 2. An example of a singular asymptotic behavior is a quadratic function $F\sim h^2$, which leads to a Big Bang singularity.

It is intriguing to notice that there is a class of $F(h)$ that grows faster than a linear function, but still, does not have finite-time singularities. An example of this is $F(h)\sim h\, \ln{h}$, or $ h\,\ln{h}\,\ln{\ln{h}},$ and so on \cite{littlerip}. These functions grow faster than a linear function, but lead to time $\tau(h)$, that depends on $h$ logarithmically, therefore, leads to nonsingular solutions. In general, if $F(h)$ can be expressed as $F(h)=g/g'$, where $g(h)$ is any function such that $g(h) \rightarrow \infty$ as $h \rightarrow \pm\infty$, then $\tau(h)$ will diverge logarithmically as $h\rightarrow \pm\infty$, which leads to a nonsingular solution. To conclude, except for $F(h)$ with a special form (as in the above mentioned cases), the solution reaches a point where $F(h)\rightarrow \pm\infty$ in a finite time if $F(h)$ grows faster than a linear function.\\
{\bf Qualitative Description:}
In this section we take the pressure $p(h)$ to be a continuous function of $h$, or at most has finite discontinuities. Therefore, we are not going to allow any infinite discontinuities for $p'(h)$, since this leads to a divergent speed of sound that violates causality.

One can qualitatively describe a solution as follows;\\
a) A solution starts from an initial value $h_0$, then, develops towards either a future fixed point or a point where $F(h) \rightarrow \pm\infty$. \\
b) If it develops towards a fixed point, then, the time taken by a solution to reach this point is always infinite according to the argument in section 2 and the solution has no future finite-time singularities.\\
c) If it develops towards a point where $F(h) \rightarrow \pm\infty$, and the asymptotic behavior is linear or slower, then, the time to reach this point is infinite and the solution has no future finite-time singularities.\\
d) If it develops towards a point where $F(h) \rightarrow \pm\infty$, and the asymptotic behavior is growing faster than a linear function but $F(h)$ has the asymptotic form $F(h)\sim g/g'$, where $g(h)$ is any function such that $g(h) \rightarrow \infty$ as $h \rightarrow \pm\infty$, then, the time to reach this point is infinite and the solution has no future finite-time singularities.\\
e) If it develops towards a point where $F(h) \rightarrow \pm\infty$, and the asymptotic behavior is growing faster than a linear function and $F(h)$ has the asymptotic form $F(h)\sim g/g'$, but $g(h)$ does not grow as $g(h) \rightarrow \infty$ when $h \rightarrow \pm\infty$, then, the time to reach this point is finite and the solution is singular.\\
f) In all the above cases the solution is a monotonic function of time as it develops from $h_0$ towards either a fixed point or a point where $F(h) \rightarrow \pm\infty$.\\
The same analysis can be followed backward in time but with past fixed points and points where $F(h) \rightarrow \pm\infty$. The solution has no finite time singularities if it has no future and no past time singularities.

\section{Examples and Application}
In this section we use the qualitative method developed in the previous sections to describe the general behavior of viscous fluids in FLRW models and compare it with exact solutions. The last exact solution with $r=1/4$ describes either as a nonsingular phantom fluid or a unified dark fluid. At the end of this section we use the phase space method, causality and stability constraints to list possible future scenarios of the universe.

\subsection{Examples: Nonsingular viscous fluids}

Consider the following equation of state (EoS)
\be p(H)=(\gamma-1)\,\rho-3\,\eta(\rho)\,H,\ee
where $\eta(\rho)=\eta_0\, \rho^r$. The above EoS can describe a fluid with bulk viscosity $\eta(\rho)$ (see e.g., \cite{cas_thermo_rel}), a polytropic fluid \cite{peerie-henri}, or a fluid with adiabatic particle production (see e.g., \cite{lima}). Although these different interpretations of the above EoS produce the same dynamics, their thermodynamics could be different. Several solutions for the above EoS with different values of $r$, including the cases discussed here, are known in the literature, see for example \cite{barrow2}. Here we are going to discuss two viscous solutions, the one with $r=1$ which is well known in the literature \cite{murphy} and another with $r=1/4$, which is less known and express them in terms of Lambert-W function. This clearly shows how the density and the scale factor behave as functions of time.

Now using the above pressure in Eqn. \ref{H} we obtain
\be \dot{H}= -{1\over 2}\,(\rho+p)=-{3\over 2}\,(\gamma\,H^2-3^r\,\eta_0\,H^{2r+1})\label{visco}\ee
Taking $h=H/H^*$ and $\tau=c\,H^*\,t$, where $H^*=({\gamma \over 3^{r}\,\eta_0})^{1 \over 2r-1}$ and $c={3\,\gamma / 2}$ we get
\be {dh \over d\tau} =-h^2\,(1-h^{2r-1}), \hspace{.3in} c\, {1 \over a}\,{da \over d\tau}=h,\ee
Notice that we have two fixed points, $h_{1,2}=0,1$, for this equation. The nature of these fixed points depends on the value of $r$. Solutions with $r>1/2$ behave differently compared to those with $r < 1/2$. In the first case $F(h)$ is negative during its interpolation between the two fixed points, while the reverse is true for the second case. It is interesting to notice that the first case describes a nonsingular universe which has an EoS parameter $w(\rho)\geq-1$ since $F(h)=-1/2\,(\rho+p)$ while the second case describes a nonsingular universe dominated by a phantom component with EoS parameter $w(\rho)\leq-1$.
\begin{figure}[htp]
 \centering
  \includegraphics[angle=-90,width=90mm]{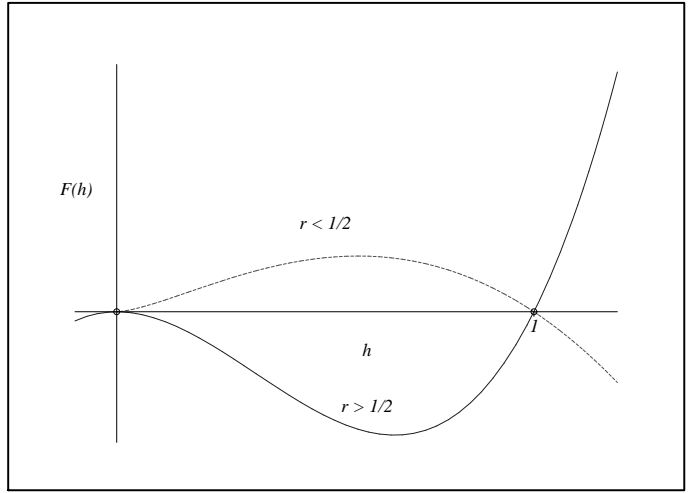}
 \caption{\footnotesize Fluids with $r>1/2$ and $r<1/2$}
 \label{fig-4}
 \end{figure}
It is interesting to notice that fluids with $r<1/2$ and $h_0>1$ have a similar behavior to that of a generalized Chaplygin-gas cosmology, i.e., for large scale factor, it behaves as a cosmological constant and for small scale factor it behaves as a fluid with EoS $p=(\gamma-1) \rho$. One can see that by writing Eqn.(\ref{one}) as
\be a\,{d\rho(a) \over da}=-3\,(\rho(a)+p(a))=-3\,\gamma\,\rho\,(1-\eta'\,\rho^{r-1/2}),\ee
where $\eta' =\sqrt{3}\, \eta_0$ solving the above equation for a general viscous fluid with $r-1/2=-s<0$, we get

\be \rho(a)=(\gamma/\eta')^{-1/s}\,\left[ 1+C_1\,a^{-3\gamma s}\right]^{1/s}\ee
If $\rho_0> (\eta'/\gamma)^{1/s}=\rho^*$, the integration constant $C_1$ is positive, sice
\be C_1= a_0^{3\gamma s}\,\left[\left({\rho_0 / \rho^*}\right)^{s}-1\right],\ee then for small $a$, i.e., $(a_0/a)^{3\gamma s}[\left({\rho_0 / \rho^*}\right)^{s}-1]>> 1$, we get
\be  \rho\sim a^{-3\gamma}\ee  which describe a fluid with an EoS $p=(\gamma-1) \, \rho$, and for large $a$ we get
\be  \rho\sim \rho^* \ee
which describes an empty space with a cosmological constant $\rho^*$.
Another interesting feature of these models is that they show that a normal matter, i.e.,$\gamma > 0$, with bulk viscosity, behaves as a phantom matter. Furthermore, in the $r<1/2$ case these solutions are nonsingular. To show these features, let us start with the pressure \be p=\gamma \, \rho \left( 1-(H^*/H)^{s}\right)-\rho, \ee which leads to an effective EoS parameter
\be w_{eff}=\gamma \, \left( 1-(H^*/H)^{s}\right)-1,\ee
since $\gamma>0$, and $H^*/H>1$, we always have $w_{eff}\leq -1$, which breaks all energy conditions. Since $H(t)$, in this case, interpolates between two fixed points ($H^*$ and $0$) and the pressure $p(H)$ is continuous and differentiable, then according to the argument in section 2, the solution is nonsingular and takes an infinite time to reach a fixed point. We are going to see an explicit example of this behavior for the $r=1/4$ case.

\begin{figure}[htp]
 \centering
  {\includegraphics[angle=-90,width=80mm]{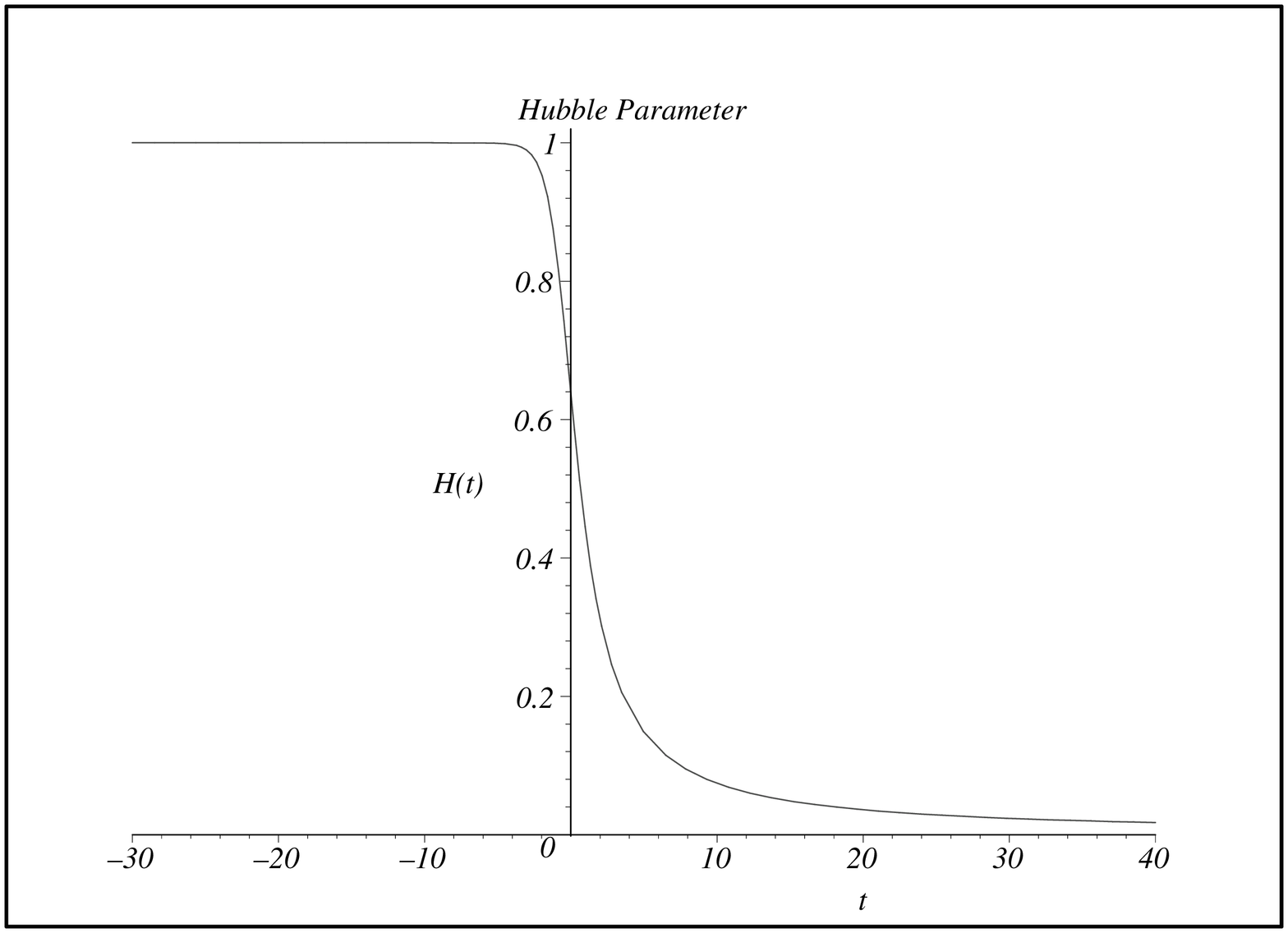}}{ \includegraphics[angle=-90,width=80mm]{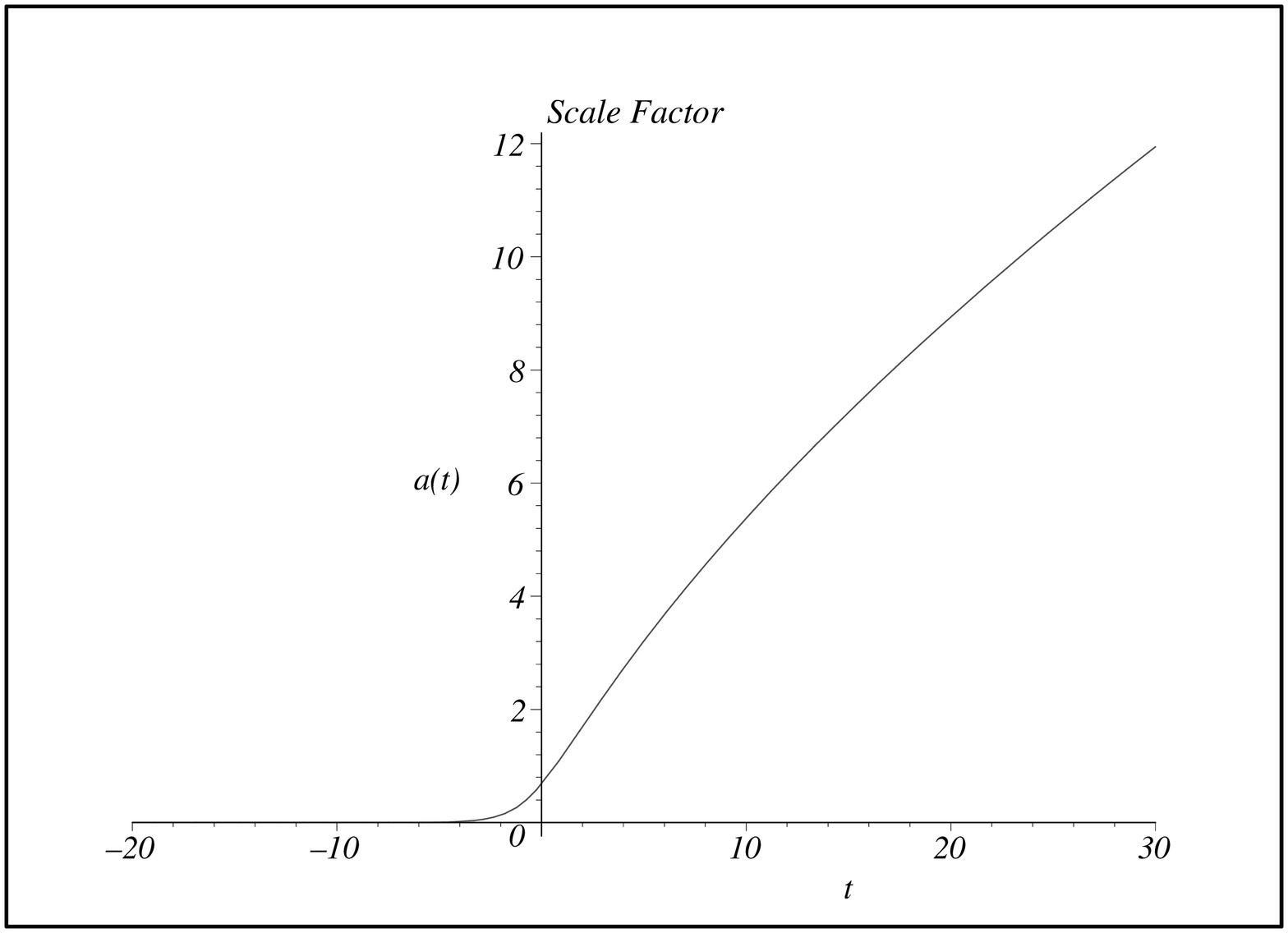}}
 \caption{\footnotesize Hubble parameter $H(t)$, in units of H$^*$, and scale factor $a(t)$, in units of $a_0$, versus time $t$ in units of $1/H^*$, $r=1$ and $\gamma=1$.}
 \label{fig-5}
 \end{figure}

{\bf Fluids with r=1 :}
 This case was first discussed in \cite{murphy} as nonsingular solution for a viscous fluid cosmology. In literature this solution usually expressed in terms of time as a function of the scale factor or the density. Here we express the energy density and the scale factor as functions of time in terms of Lambert W-function. First, let us analyze this case qualitatively taking subsection 3.2 in consideration. The asymptotic behavior has the form $F(h)\sim h^3$ which leads to singular solutions unless there are fixed points. $F(h)$ has two fixed points $h_1=0$ and $h_2=1$, as shown in Figure (\ref{fig-4}). The first is half-stable point and the second is an unstable point. These points divide possible solutions into three types; i) a solution where $h\in (-\infty,0]$, if $h_0<h_1$, ii) a solution where $h\in[0,1]$, if $h_2>h_0>h_1$, and iii) a solution where $h\in[1,\infty)$, if $h_2<h_0$. Notice that, in case b), since $p(h)$ is differentiable then by the argument in section 2, it takes the solution an infinite time to reach $h_1$ starting from some initial value, $h_0$, where $h_1<h_0<h_2$. The same is true if we calculate the time taken by the solution to go from $h_2$ to $h_0$. Therefore, for $h_1<h_0<h_2$ the solution is nonsingular and interpolates smoothly between $h_1$, and $h_2$. If $h_0<h_1$ or $h_0>h_2$ the solution has a finite time singularity since the asymptotic behavior of $F(h)\sim h^3$ is growing faster than a linear behavior. For $h_0>h_2$ the solution describes a universe that starts from a de Sitter space and endes with a Big Rip singularity.
\begin{figure}[htp]
 \centering
  \includegraphics[angle=-90,width=90mm]{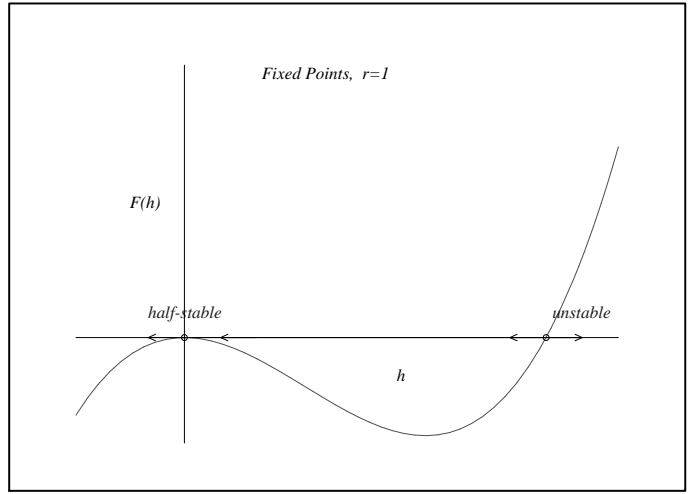}
 \caption{\footnotesize Fixed points for $r=1$ case}
 \label{fig-4}
 \end{figure}
The above equations can be solved exactly in terms of Lambert W-function, which is the solution of the equation $W\,e^{W}=x$. The hubble parameter and the scale factor in terms of the time "t" are given by
\bea H(t)&=&H^* \,[W(c_1\, e^{(3\,\gamma/2)\,H^*\,t})+1]^{-1}, \nonumber\\
a(t)&=&c_2\,[W(c_1\, e^{(3\,\gamma/2)\,H^* \,t})]^{2/3\, \gamma}\eea
Having $a(t_0)=a_0$, and $\rho(t_0)=3\,{H_0}^2$ at $t=t_0$, we get
\be \rho(t)=3\,{H^*}^2 \,[W(\beta\, e^{(3\,\gamma/2)\,H^*\,(t-t_0)})+1]^{-2}, \ee \be a(t)=a_0\,\left[{W(\beta\, e^{(3\,\gamma/2)\,H^*\,(t-t_0)}) \over W(\beta\, )}\right]^{(2/3\, \gamma)}\ee
where $\beta$ is given by
\be \beta = \left({H^* \over H_0}-1 \right)\,e^{({H^* \over H_0}-1)}\ee
It is crucial at this point to know the sign of $\beta$ since it controls the behavior of the W-function. The $h_0=H_0/H^*<1$ initial value corresponds to a positive $\beta$, which leads to a smooth behavior in all times for the energy density. For the scale factor, the moments at which it either diverges or goes to zero are when $t=+\infty$ and $t=-\infty$ respectively. Therefore, we have no finite-time singularities in this case. In early times, this model describes an empty universe with a cosmological constant $\Lambda\sim {H^*}^2$, which evolves to a universe with an EoS $p=(\gamma-1)\,\rho$ in late times.
The behavior of $H(t)$ and $a(t)$ as a function of time $t$, is shown in Figure(\ref{fig-5}), which is clearly monotonic.
If the initial value $h_0=H_0/H^*>1$ ($\beta <0$ in the case), then we have a singularity of type-I, or a Big Rip singularity. In this case the effective EoS parameter $w_{eff}=p/\rho <-1$, therefore the fluid is phantom. Notice that, if $h_0<h_2$, the $r<1/2$ cases describe nonsingular phantom solutions.

One can modify the above model to accommodate the late-time acceleration by adding a cosmological constant $\Lambda$. The exact solution of this modified model is not easy to get, but if $\Lambda/{H^*}^2=\lambda<<1$, or the cosmological constant of the early times is larger than that of late times, then the fixed points become
\be h_1\sim \sqrt{\lambda}, \hspace{.3in} h_2\sim 1-\lambda,\ee
This model describes a universe that interpolates between two de Sitter spaces one with large cosmological constant in early times and another with small cosmological constant in late times which can model the inflation and late time acceleration periods.\\
{\bf  Fluid with r={1/4}:}
Here we discuss the solution for the $r=1/4$ case, which can be expressed in terms of Lambert W-function. As we mentioned above, cases with $r<1/2$ are interesting since they have a unified dark fluid behavior when $h_0>h_2$. If $h_0<h_2$, the solution describes a nonsingular phantom matter with $w_{eff}\leq -1$.
To analyze this case qualitatively, let us start with the asymptotic behavior of $F(h)$ which has the form $F(h)\sim h^2$. It clearly leads to singular solutions unless we have fixed points. $F(h)$ has two fixed points $h_1=0$ and $h_2=1$, as shown in Figure (\ref{fig-6}). The first is an unstable point and the second is a stable point. These points divide possible solutions into two types; i) a solution where $h\in[0,1]$, if $h_2>h_0>h_1$, and ii) a solution where $h\in[1,\infty)$, if $h_2<h_0$. Notice that, in the first case, i), since $p(h)$ is differentiable then by the argument in section 2, it takes the solution an infinite time to reach $h_1$ starting from some initial value, $h_0$, where $h_1<h_0<h_2$. The same is true if we calculate the time to go from $h_2$ to $h_0$, therefore, the solution is nonsingular and interpolates smoothly between $h_1$, and $h_2$. The second case ii) is singular, since the asymptotic behavior $F(h)\sim h^2$ is growing faster than a linear behavior. In fact, this solution describes a universe that starts from a finite-time singularity in the past (Big Bang type) and evolves to a de Sitter space after an infinite time.
\begin{figure}[htp]
 \centering
  \includegraphics[angle=-90,width=80mm]{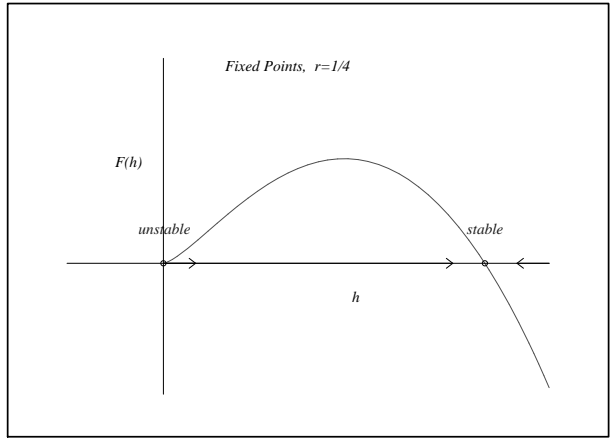}
 \caption{\footnotesize Fixed points for $r=1/4$ case}
 \label{fig-6}
 \end{figure}

\begin{figure}[htp]
 \centering
  {\includegraphics[angle=-90,width=80mm]{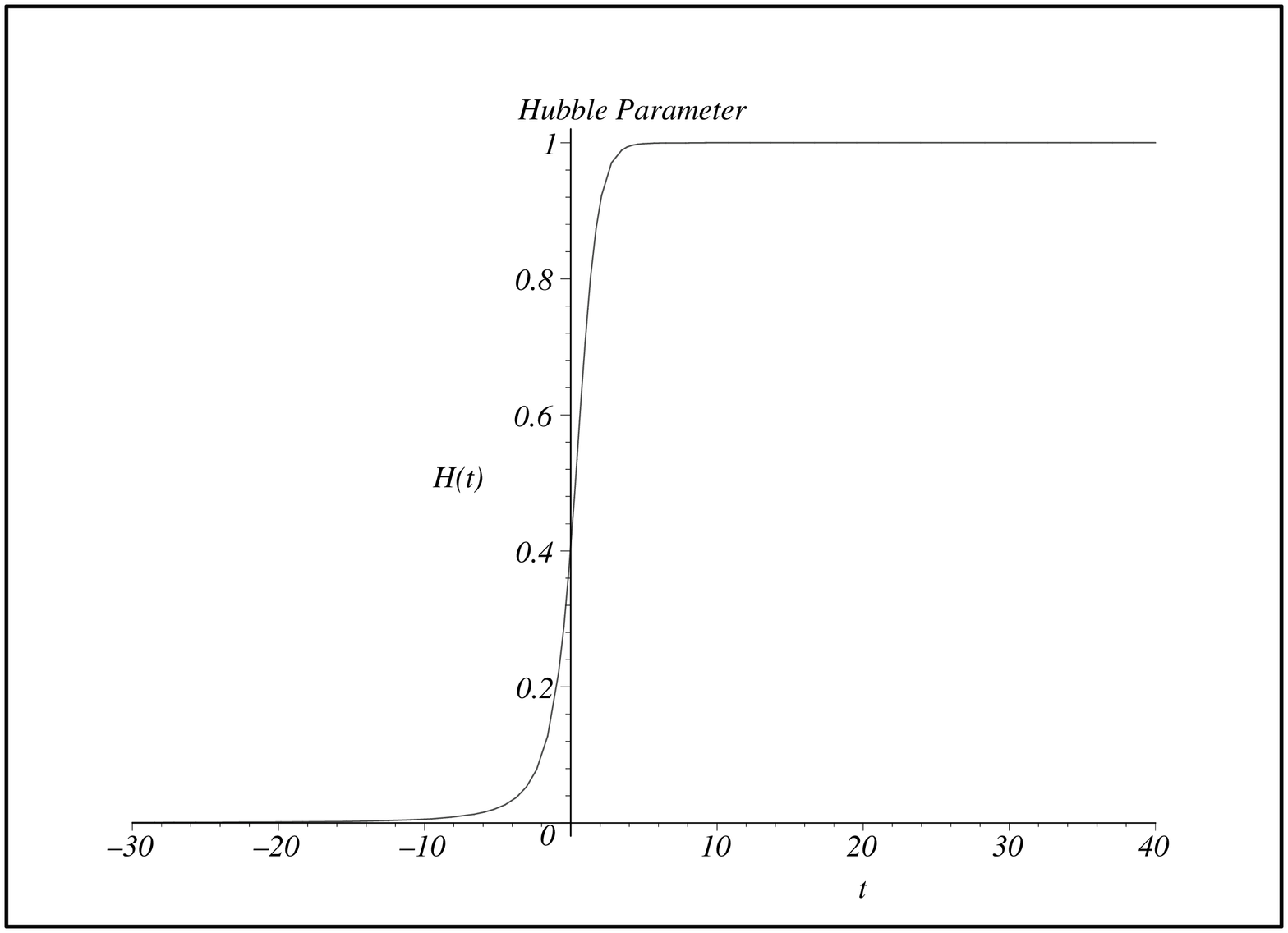}}{ \includegraphics[angle=-90,width=80mm]{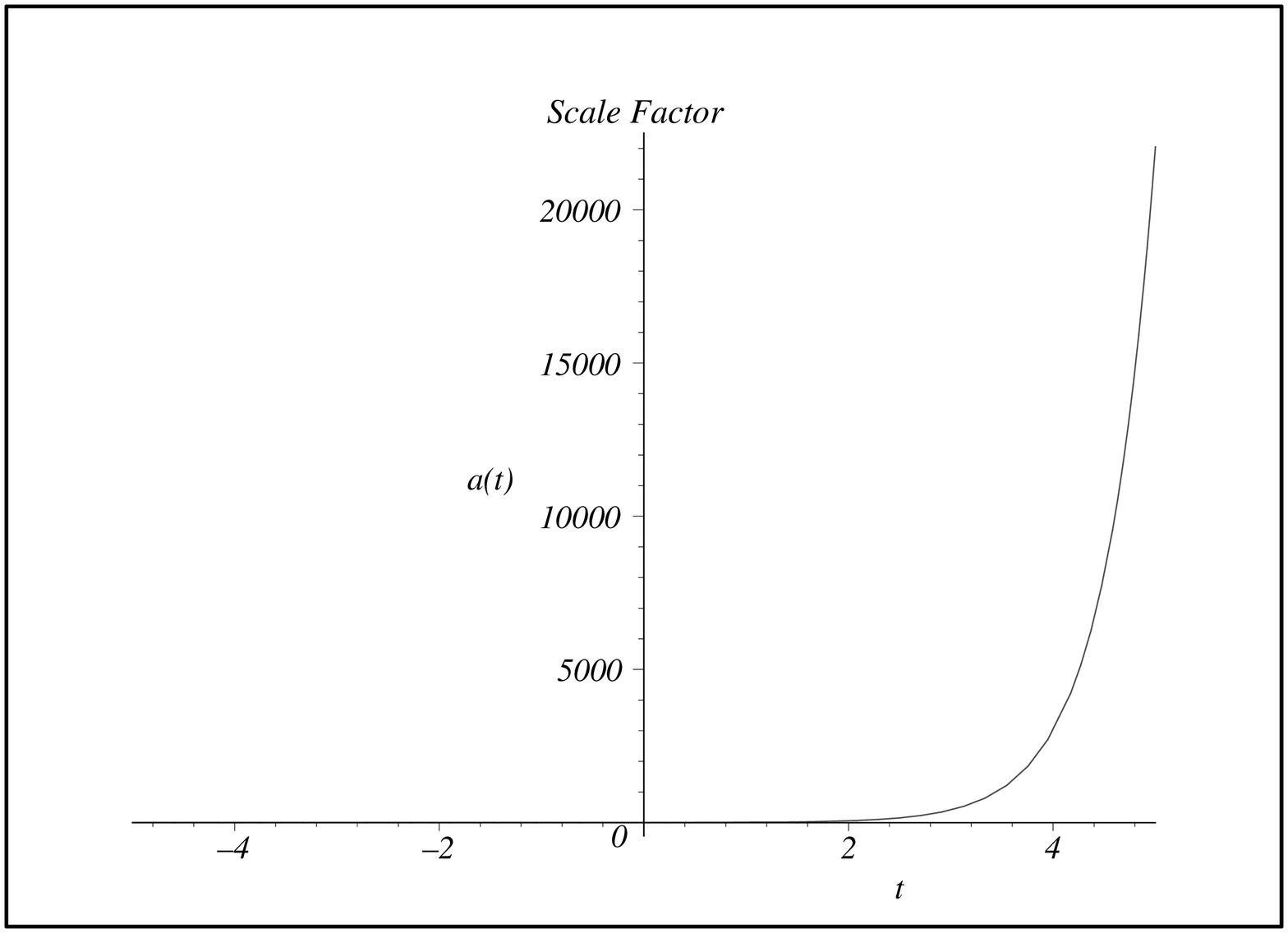}}
 \caption{\footnotesize Hubble parameter $H(t)$, in units of H$^*$, and scale factor $a(t)$, in units of $a_0$, versus time $t$ in units of $1/H^*$, $r=1/4$ and $\gamma=1$.}
 \label{fig-8}
 \end{figure}

Eqn.(\ref{visco}) for $r=1/4$ case has an exact solution which has the following form
\bea H(t)&=&H^* \,[W(c_1\, e^{(-3\,\gamma/4)\,H^*\,t})+1]^{-2}, \nonumber\\
a(t)&=&c_2\,\left[{W(c_1\, e^{(-3\,\gamma/4)\,H^* \,t})+1 \over W(c_1\, e^{(-3\,\gamma/4)\,H^* \,t})}\right]^{4/3\, \gamma}. \eea Using initial conditions, the integration constants are
\be c_1=e^{{3\, \gamma \over 4}\,H^*\,t_0}\,\beta', \hspace{.3in} c_2=a_0\, \left[{W(\beta')+1 \over W(\beta')}\right]^{4 \over 3\,\gamma},\ee where \be\beta'=\left(\sqrt{H^*/H_0}-1\right)\,e^{\left(\sqrt{H^*/H_0}-1\right)}.\ee
Notice that $\beta'$ is positive for $h_0<1$ and negative for $h_0>1$. For initial value $h_0<1$ one expects a nonsingular solution which interpolates monotonically between Minkowski space and de Sitter space. The Hubble parameter $H(t)$ and the scale factor $a(t)$, as functions of time "t", are shown in Figure(\ref{fig-8}).
 As we mentioned in the beginning of the section, this class of solutions shows how a viscous fluid with a usual EoS (i.e., $\gamma >0$) behaves as phantom component. Similar to the $r=1$, one can consider adding a small cosmological constant. A small cosmological constant $\Lambda/{H^*}^2=\lambda<<1$ leads to the following fixed points \be h_1\sim {\lambda}^{2/3}, \hspace{.3in} h_2\sim 1-2\,\lambda.\ee The solution with initial value $h_1<h_0<h_2$ described a universe filled with a phantom matter interpolating between a de Sitter space with a small cosmological constant at early times and another with a large cosmological constant in late times.

\subsection{Fate of the Universe}
Here we list all possible future scenarios of the universe as a single fluid in FLRW cosmology without assuming any fixed points but imposing the causality and stability constraints. It is known that, if $p>-1/3 \rho$, destiny of the universe is tied to geometry (see for example \cite{linder-review}) and the value of $k$ is important to predict the fate of the universe. On the other hand, if $p<-1/3 \, \rho$ destiny is not tied to geometry but controlled by the behavior of the energy density $\rho$. This can be shown if we consider the known mechanical model for $a$, by rewriting Friendmann equation as \be \dot{a}^2={1 \over 3}\,a^2\, \rho(a)-k =-2\,V_{eff}(a).\ee
Taking the EoS $p=w\,\rho$ leads to
\be \rho(a)=C\, a^{-3\,(1+w)}\Rightarrow 2\,V_{eff}(a)=-C/3\, a^{-\,(1+3w)}+k\ee
For large $a$, if $w>-1/3$, $k$ controls the existence of vanishing velocities, or turning points, but for $w>-1/3$, the potential gets a small contribution from $k$ compared to that coming from $a^2\rho$. This breaks the connection between the geometry and destiny.
In fact, this simple argument is also suggesting that our universe will keep on expanding because of the domination of the dark energy component. We will see next that this is not generally the case.\\
Here we are going to use the above constraints to list possible scenarios for the future of the universe. Let us model our universe using a general single barotropic fluid, which is a reasonable assumption since dark energy is dominating in late times. \be \dot{H}= 1/2 \, (3\,H^2+p(H)).\ee
First, it is easy to show that we are in a region in the phase space (i.e., $\dot{H}-H$ space ) where $\dot{H}<0$\footnote{One can directly calculate $\dot{H}$ using the best measured values for $w_{D}$, $\Omega_m$ and $\Omega_D$.}. It is known that the deceleration parameter has changed sign from positive to negative as the universe evolved from a matter dominating era to a dark energy dominating era. To see how this crossing happened consider first the zero acceleration curve which is given by
\be {\ddot{a} \over a}=\dot{H}+H^2=0 \hspace{.3in} \Rightarrow \hspace{.3in}\dot{H}=-H^2,\ee then, as an example of dark energy, consider an accelerating universe with a cosmological constant and matter
\be \dot{H}=F(H)=-1/2\, (3 H^2-\rho_{\Lambda}).\ee
By plotting both functions $-H^2$ and $-1/2\, (3 H^2-\rho_{\Lambda})$ in Figure(\ref{fig-6}) one can see how this crossing happened, i.e., going from $\ddot{a}<0$ to $\ddot{a}>0$.
\begin{figure}[htp]
 \centering
  {\includegraphics[angle=-90,width=80mm]{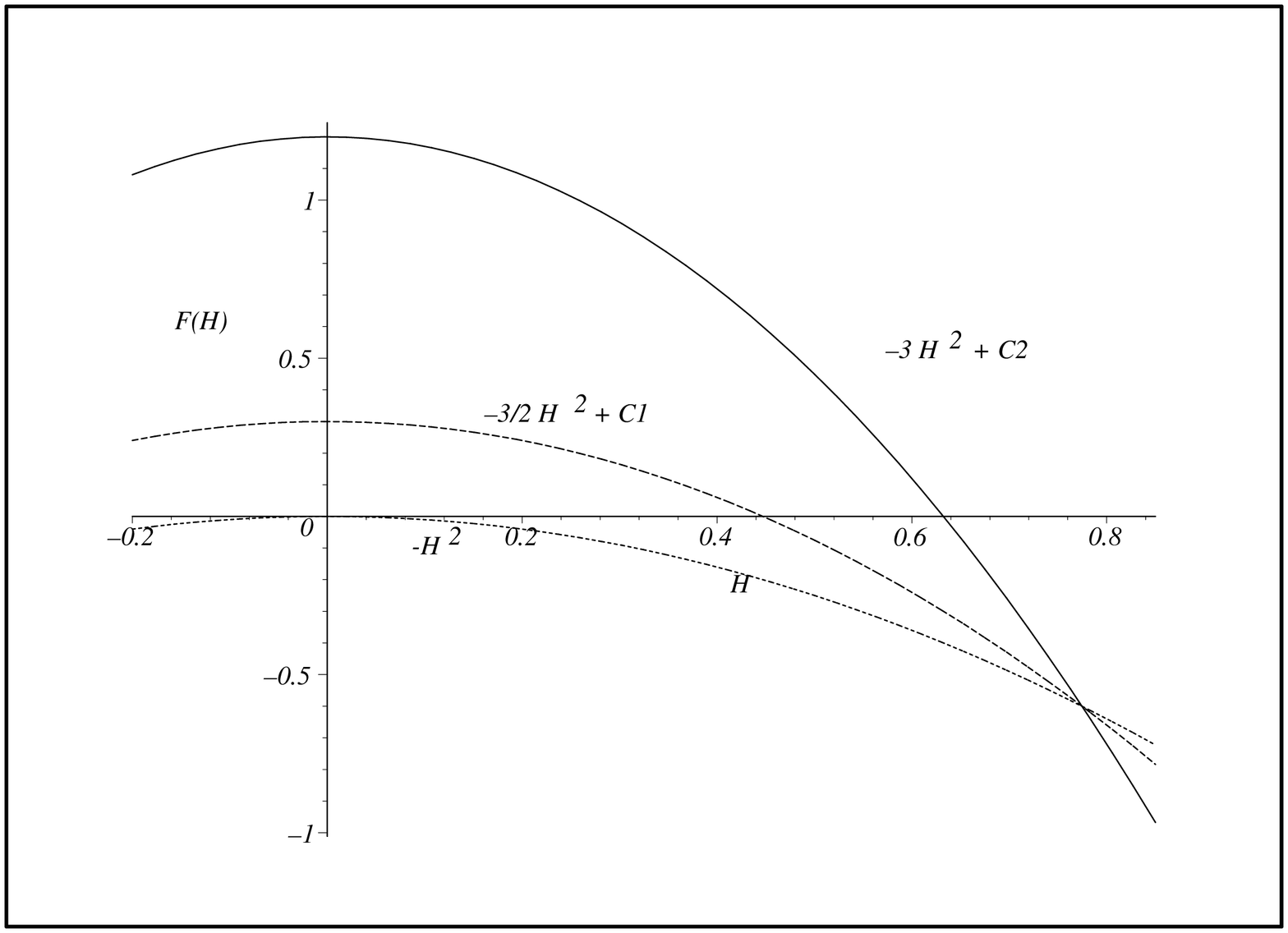}\includegraphics[angle=-90,width=80mm]{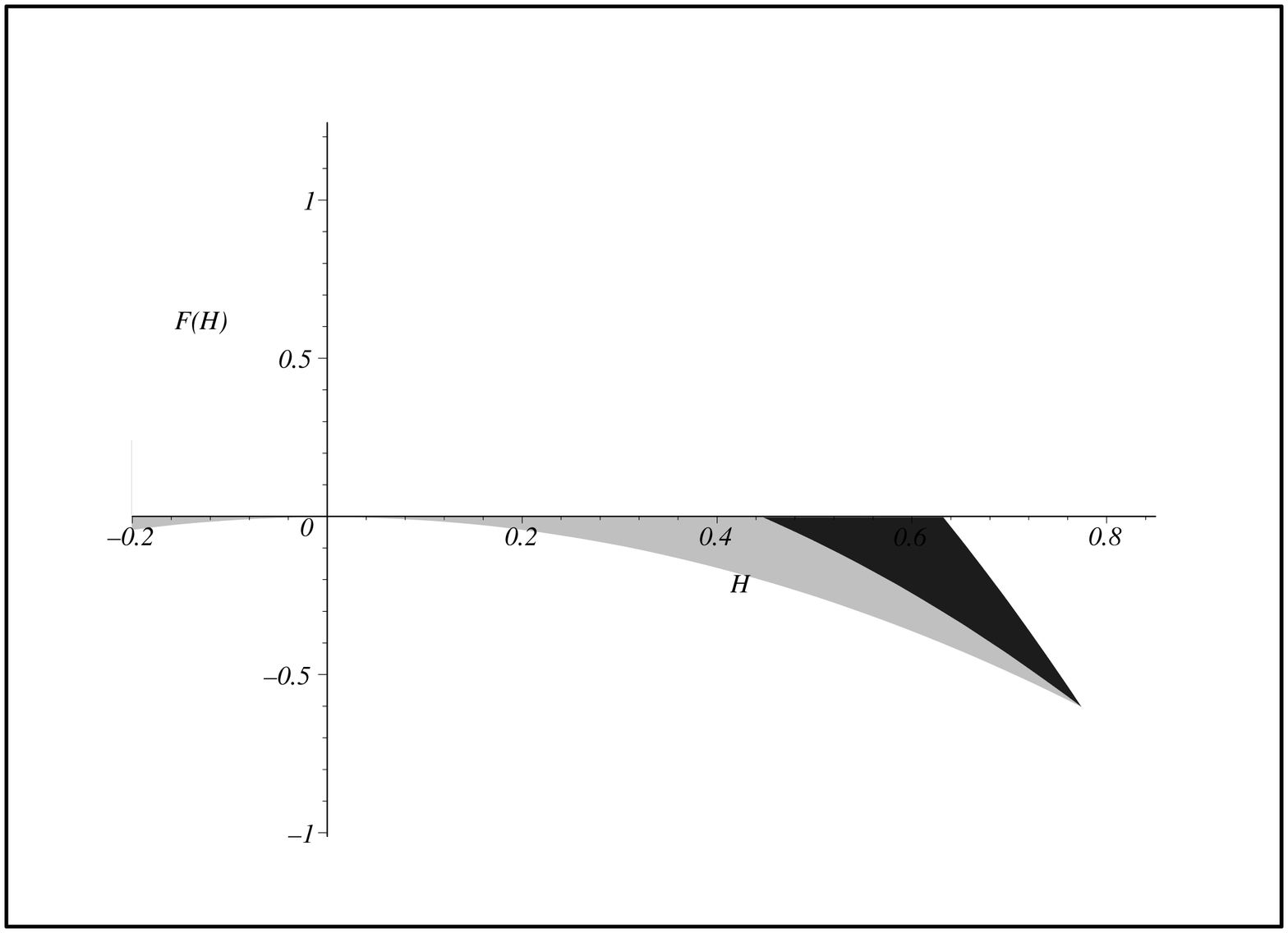}}
 \caption{\footnotesize The $-H^2$ curve is the zero-acceleration curve. The black region between the two curves, $-3/2\,H^2$ and $-3/2\,H^2$ satisfies the two constraints on barotropic fluids and any curve in this region must end with a fixed point on positive $H$-axes}
 \label{fig-6}
 \end{figure}
 The point at which the acceleration vanishes can be used as a reference point to draw the constraints on possible evolutions of the universe. Using the causality and stability constraints, $dp/d\rho \leq 1$ and $ dp/d\rho\geq 0$, we get $0 \geq dp(H)/dH \leq 6\,H$ which in turn leads to $-3\,H\geq dF(H)/dH \leq -6\,H$. Integrating this inequality leads to $-3/2\,H^2 +C_1 \geq F(H) \leq -3\,H^2 +C_2$, where the integration constants $C_1$ and $C_2$ are fixed by the initial values of $H$ and $\dot{H}$ at the reference point. Notice that, $C_1$ and $C_2$ have to be positive numbers, otherwise crossing the zero-acceleration curve will not occur. The last inequality constrains the future behavior of $F(H)$ to lie between these two parabolas, as a result, $F(H) $ must meet the $H$-axes in a future time and ends as a de Sitter universe after infinite time. In addition, this evolution starting from the point of zero-acceleration till the end of time is nonsingular since it ends with de Sitter universes.

As one might notice, although, the causality constraint is essential for any physical model. It is not clear if we should insist on having the stability constraint, since we do not know the physics of the dark energy component. Now, if we relax the stability constraint, $ dp/d\rho\geq 0$ and allow the pressure derivative to go negative. From the above phase space diagram, it is clear that the universe either ends as an empty universe or hits the negative $\dot{H}$-axes. The last possibility is interesting since according to the discussion in subsection 3.1 it describes a turnaround behavior, therefore, the universe in a future finite time reaches a maximum size, then, recollapses, since the hubble parameter $H$ changes sign.

\section{Conclusion}

In this work, we used a phase space method to study possible consequences of having fixed points in a single fluid flat FLRW models. Some of these are; {\it (i)} if we describe our universe as a single component fluid with a future fixed point, then the resulting cosmology does not have future-time singularities of types I, II and III in \cite{no-od-class}, {\it (ii)} cosmologies with a future and a past fixed points are free of of types I, II and III singularities, {\it (iii)} one can use a simple argument to show the phantom divide \cite{Hu-2005,Knuz-2006}, or in a single fluid FLRW models it is impossible for a physical solution to cross the phantom divide line in a finite time, and {\it (iv)} in these models, the only way to get bounce solutions is to have a nonvanishing pressure as $\rho \rightarrow 0$. This method can be used to construct nonsingular late-time models, in particular, unified dark fluid and dark energy models. We use this method to qualitatively describe any flat FLRW model with fixed points. We discussed FLRW cosmology with bulk viscosity $\eta \sim \rho^r$, and presented two exact solutions with $r=1$ and $r=1/4$, which are expressed in terms of Lambert-W function. The last solution describes either a nonsingular phantom dark energy or a unified dark fluid model. The phantom solution is interesting since it shows how a viscous normal fluid behaves very similar to a phantom matter without Big Rip singularities. In addition, it interpolates between two de Sitter spaces with small and large cosmological constants. Possible future scenarios of our universe include; a de Sitter space, an empty universe with vanishing cosmological constant, or a turn a round solution that reaches a maximum size, then collapses.\\
{\bf Acknowledgement}\\
I would like to thank P. Argyres, S. Das, A. Shapere, E. Lashin and A. El-Zant for several discussions and comments.


{\small
\begin{thebibliography}{99}
\bibitem{Riess}
A. G. Riess et al. [Supernova Search Team Collaboration], Astrophys. J. 607, 665 (2004)
[astro-ph/0402512];
R. A. Knop et al., [Supernova Cosmology Project Collaboration], Astrophys. J. 598, 102 (2003)
[astro-ph/0309368];
A. G. Riess et al. [Supernova Search Team Collaboration], Astron. J. 116, 1009 (1998)
[astro-ph/9805201];
\bibitem{Perlmutter}
S. Perlmutter et al. [Supernova Cosmology Project Collaboration], Astrophys. J. 517, 565 (1999)
[astro-ph/9812133].
\bibitem{wmap}
C. L. Bennett et al., Astrophys. J. Suppl. 148, 1 (2003) [astro-ph/0302207];
D. N. Spergel et al., Astrophys. J. Suppl. 148 175 (2003) [astro-ph/0302209].
\bibitem{SDSS}
M. Tegmark et al. [SDSS Collaboration], Phys. Rev. D 69, 103501 (2004) [astro-ph/0310723];
K. Abazajian et al. [SDSS Collaboration], astro-ph/0410239; astro-ph/0403325; astro-ph/0305492;
M. Tegmark et al. [SDSS Collaboration], Astrophys. J. 606, 702 (2004) [astro-ph/0310725].
\bibitem{Cha-Xray}
S. W. Allen, R. W. Schmidt, H. Ebeling, A. C. Fabian and L. van Speybroeck, Mon. Not. Roy. Astron.
Soc. 353, 457 (2004) [astro-ph/0405340].
\bibitem{br}
R. Caldwell, M. Kamionkowski, N. Weinberg, Phys.Rev.Lett. {bf 91} 071301 (2003). 
\bibitem{barrow1}
J. Barrow, D. Kimberly, and J. Magueijo, Class. Quant. Grav. {\bf 21} 4289 (2004). 
J. Barrow, Class. Quant. Grav. {\bf 21} 5619 (2004). 
J. Barrow, C. Tsagas, Class. Quant. Grav. {\bf 22} 1563 (2005)
\bibitem{Bouhmadi-Lopez}
M. Bouhmadi-Lopez and J. Jimenez Madrid, JCAP {\bf 0505}, 005 (2005).
M.~Bouhmadi-Lopez, P.~F.~Gonzalez-Diaz and P.~Martin-Moruno, Phys.\ Lett.\ B {\bf 659}, 1 (2008).
\bibitem{caldwell_review}
 R. Caldwell and M. Kamionkowski, Ann. Rev. Nucl. Part. Sci. {\bf 59}, 397 (2009).
 \bibitem{tuener_review06}
J. Friedman and M Turner, Ann. Rev. Astron. Astrophys. {\bf 46}, 385 (2008).
\bibitem{linder-baro}
E. V. Linder and R. J. Scherrer, Phys. Rev. D {\bf 80}, 023008 (2009).
\bibitem{psww'}
R.J. Scherrer, Phys. Rev. D {\bf 73}, 043502 (2006).
\bibitem{KMP-2001}
A.Y. Kamenshchik, U. Moschella, and V. Pasquier, Phys.
Lett. B 511, 265 (2001).
\bibitem{BTV-2002}
N. Bilic, G.B. Tupper, and R.D. Viollier, Phys. Lett. B
535, 17 (2002).
\bibitem{BBS-2002}
[24] M.C. Bento, O. Bertolami, and A.A. Sen, Phys. Rev. D
66, 043507 (2002).
\bibitem{CSN-1997}
T. Chiba, N. Sugiyama, and T. Nakamura, MNRAS 289,
L5 (1997).
\bibitem{CSN-1998}
T. Chiba, N. Sugiyama, and T. Nakamura, MNRAS 301,
72 (1998).
\bibitem{BDE-2005}
E. Babichev, V. Dokuchaev, and Yu. Eroshenko, Class.
Quant. Grav. 22, 143 (2005).
\bibitem{HN-2004}
R. Holman and S. Naidu, astro-ph/0408102.
\bibitem{AB-2006}
K.N. Ananda and M. Bruni, Phys. Rev. D 74, 023523
(2006).
\bibitem{QBB-2007}
C. Quercellini, M. Bruni, and A. Balbi, Class. Quant.
Grav., 24, 5413 (2007).
\bibitem{NO-2004}
S. Nojiri and S.D. Odintsov, Phys. Rev. D 70, 103522
(2004).
\bibitem{Kremer-2003}
G.M. Kremer, Phys. Rev. D 68, 123507 (2003).
\bibitem{Capoz-2005}
S. Capozziello et al., JCAP 0504, 005 (2005).
\bibitem{no-od-class}
S. Nojiri, S. D. Odintsov and S. Tsujikawa, Phys. Rev. D {\bf 71}, 063004 (2005).
\bibitem{strogatz}
S. H. Strogatz, {\it Nonlinear Dynamics and Chaos}, Preseus Books, 1994.
\bibitem{Hu-2005}
W. Hu, Phys. Rev. D {\bf 71}, 047301 (2005).
\bibitem{Knuz-2006}
M. Kunz and D. Sapone, Phys. Rev. D {\bf 74}, 123503 (2006).
\bibitem{quintom}
Yi-Fu Cai, Emmanuel N. Saridakis, Mohammad R. Setare, Jun-Qing Xia, Phys. Rept. 493, 1 (2010).
\bibitem{curved_1}
A. Awad, in preparation.
\bibitem{littlerip}
P. H. Frampton, K. J. Ludwick and R. J. Scherrer, Phys. Rev. D {\bf 84} 063003 (20011).
\bibitem{vis_littlerip}
I. Brevik, E. Elizalde, S. Nojiri and S. Odintsov, Phys. Rev. D {\bf 84} 103508 (2011).
\bibitem{cas_thermo_rel}
R. Maartens, {\it Casual Thermodynamics in Relativity }, astro-ph/9609119.
\bibitem{murphy}
G.L. Murphy, Phys. Rev. D {\bf 8} 4231 (1973).
\bibitem{barrow2}
J. Barrow, Phys. Lett. B  {\bf 180}, 335 (1986).
J. Barrow,  Nucl. Phys. B {\bf 310}, 743 (1988). 
\bibitem{lima}
J. A. S. Lima, F. E. Silva and R. C. Santos, Class. Quant. Grav. {\bf 25} 205006 (2008).
\bibitem{peerie-henri}
P.H. Chavanis, {\it Models of universe with a polytropic equation of state: I. The early universe}, [arXiv:1208.0797].
\bibitem{linder-review}
E. Linder, {\it Mapping the Cosmological Expansion}, Rept. Prog. Phys. {\bf 71} 056901 (2009).
\bibitem{No-Od-phantom}
S. Nojiri, S. Odintsov, Phys. Rev. D {\bf 72} 023003 (2005). 

\end{thebibliography}
}
\end{document}